\documentclass[nosumlimits]{amsart}
\usepackage{amsmath}
\usepackage{amssymb}

\newtheorem{thm}{Theorem}[section]
\newtheorem{defin}[thm]{Definition}

\newtheorem{pro}[thm]{Proposition}
\numberwithin{equation}{section}

 \newcommand{\into}{\rightarrow}

 \newcommand{\set}[1]{\left\{#1\right\}}
 
 \newcommand{\dotp}[2]{\left<#1,#2\right>}

 \newcommand{\qtext}[1]{\quad\text{#1}\quad}
 \newcommand{\fa}{\qtext{for all}}
 \newcommand{\bb}{\begin{equation*}}
 \newcommand{\ee}{\end{equation*}}
 \newcommand{\bp}{\begin{proof}}
 \newcommand{\ep}{\end{proof}}

\begin{document}

\title[Solution to satellite orbit anomaly problem]
{NASA's satellite orbit anomaly problem\\ can be solved precisely\\
in the frame of Einstein's special theory of relativity.\\
Anomaly confirms that gravity fields propagate with velocity of
light as Einstein predicted}
\author{ Victor M. Bogdan }

\address{Department of Mathematics, McMahon Hall 207, CUA, Washington DC 20064, USA}

\date{12 October 2009}

\email{bogdan@cua.edu}


\begin{abstract}
NASA's Jet Propulsion Laboratory put on You Tube a problem that
has been baffling the scientists for sometime. It involves an
unexpected force acting on the space probes.

The author proves that NASA'S satellite orbit anomaly problem can
be solved in the frame of Einstein's Special Theory of Relativity.
The anomaly confirms that gravity fields propagate with velocity
of light as Einstein predicted.

The proof is based on the authors discovery of the relativistic
version of Newton's gravity field. The author provides formulas
for relativistic equation of motion for a spacecraft in the joint
gravitational field of the Earth and the Sun in a Lorentzian frame
attached to the Earth. The formulas are suitable for digital
computers and can be easily implemented. He also shows how to find
solutions of the relativistic equations of motion for the
spacecraft.
\end{abstract}

\bigskip

\maketitle

\bigskip

\section{NASA is baffled by an unexpected force\\
 acting on space probe}

The following is a transcript from You Tube video apparently made
by Mr. Anderson from NASA's Jet Propulsion Laboratory
\cite{anderson}:

{\em
 Mysteriously five space craft that flew past the earth have
each displayed unexpected anomalies in their motions. {\bf Pioneer
anomaly} has hints that unexpected forces may appear to act on
spacecraft. The anomalies were seen with identical Pioneer 10 and
11 spacecraft. Both seem to experience tiny but unexplained
constant acceleration toward the sun.

In five of the six flybys, the scientists have confirmed
anomalies. "I'm feeling both humbled and perplexed" said Anderson.
"There is something strange going on with spacecraft, with respect
to Earth's equator. It suggests that the anomaly is related to
Earth's rotation"

"Another thing in common between the Pioneer and these flybys is
what you would call an {\bf unbound orbit around a central body,}"
Anderson said. "For instance, the Pioneers are flying out of the
system. There's definitely something going on."

"Could Earth's rotation be a part of the equation? Rotation
forecasts are important to NASA's Jet Propulsion Laboratory."

"The more forceful winds double the angular momentum of the
atmosphere. Winds can effect the Earth's rotation."

"Something's effecting the satellites orbit. Could it be Planet X?
What do you think?"}

The purpose of this note is to answer the questions rased in this
video and to provide the gravity field formulas in the frame of
Einstein's special theory of relativity. These formulas can be
used for navigation in gravitational fields in the presence of
several celestial bodies in motion.
\bigskip

\section{The solution of the problem}
\bigskip

The answer comes from Einstein's special theory of relativity
involving Maxwell's equations and formulas discovered by the
author for relativistic gravity field that correct the Newton's
formula. The anomaly indeed is due to Earth's rotation.

With respect to the Lorentzian frame attached to Earth the Sun
moves around the Earth almost periodically with period of 24
hours. The distance to the Sun is approximately
\begin{equation*}
 r=8\,[min]\times c\,[km/sec]
\end{equation*}
so the circumference of Sun's   orbit around the Earth in the
Lorentzian frame is $2\pi r \cos(\alpha),$ where $\alpha$ is the
angle between the plane of Earth's equator and the plane of Sun's
orbit,
 and the speed of the Sun in that frame is
 approximately $\frac{1}{30}\, \cos(\alpha)\,c.$

This fraction $$q=\frac{1}{30}\cos(\alpha)$$ represents an
important factor. It determines the rate of convergence in the
computation of the field representing the retarded time.

For the sake of simplicity select the unit of time so that the
speed of light in the Lorentzian frame is $c=1.$ Select also the
unit of mass so that gravitational constant $G=1.$

This means that if the unit of length is 1 meter, then the unit of
time is approximately 3.3 nanoseconds and the unit of rest mass is
about 3.8 metric tons.

Assume that we have the formula $t\mapsto r(t)\in R^3$ for the
trajectory of the Sun around the Earth in our Lorentzian frame. It
is almost periodic. So we may a priori estimate a bound on its
velocity $w(t)=\dot{r}(t)$ in that frame. In our case it is $q.$

Let the pair $(y,t)\in R^3\times R$ denote any point in the
Lorentzian frame. According to Einstein's theory of relativity
\cite{einstein2a} and his work jointly with Rosen \cite{einstein4}
on gravity, gravity waves propagate through space with velocity of
light.

For a gravity wave from the Sun to arrive at a point $y\in R^3$ at
time $t,$ the wave should be emitted from position $r(\tau)$ on
its trajectory at some earlier time $\tau.$ This leads to the
Lorentz relation
\begin{equation*}
 t-\tau=|y-r(\tau)|
\end{equation*}
Since the Sun's trajectory is fixed the above equality implicitly
defines the function $(y,t)\mapsto \tau.$ Notice that for given
$(y,t)$ the value $\tau$ is a fixed point of the function
$f(s)=t-|y-r(s)|$ considered on the interval $s\le t.$ The
function $f$ is a contraction and its Lipschitz constant is $q.$
From Banach's fixed-point theorem, we get a fast converging
algorithm for computing the function $\tau(y,t).$ Namely use the
recursive formula
\begin{equation*}
 \tau_0(y,t)=0\qtext{and}\tau_n(y,t)
 =f(\tau_{n-1}(y,t)\fa (y,t),\ n=1,2,\ldots
\end{equation*}

After $n$ steps we get the following approximation
\begin{equation*}
 |\tau(y,t)-\tau_n(y,t)|\le
 \frac{q^n}{1-q}|t-|y-r(0)|\,|\fa (y,t)\in R^4,\ n=1,2,\ldots
\end{equation*}

The function $\tau$ is continuous as follows from results of
Bogdan \cite{bogdan65}, Theorem 3.1. Introduce the scalar field
$T$ called the time delay field by the formula
\begin{equation*}
 T(y,t)=t-\tau(y,t)\fa (y,t)\in R^4
\end{equation*}
and vector field
\begin{equation*}
 r_{12}(y,t)=y-r(\tau(y,t))\fa (y,t)\in R^4
\end{equation*}
representing a vector starting on the trajectory of the Sun at
retarded time and having its end at the point $y\in R^3$ at time
$t.$ It replaces Newton's radius vector. Notice the equality
$$T(y,t)=|r_{12}(y,t)|$$ for all $(y,t)\in R^4.$

Introduce also the velocity $v$ and acceleration $a$ vector fields
by the formulas
\begin{equation*}
 \begin{split}
    v(y,t)&=w(\tau(y,t))\\
    a(y,t)&=\dot{w}(\tau(y,t))
 \end{split}
\end{equation*}
for all $(y,t)\in R^4.$

On the set
\begin{equation*}
 G=\set{(y,t)\in R^4:\ T(y,t)>0}
\end{equation*}
introduce the scalar field $u=1/T,$  the unit vector field
$e=ur_{12},$ representing direction of vector $r_{12},$ and the
scalar field
\begin{equation*}
 z=\frac{1}{1-e\cdot v}.
\end{equation*}
Notice that since the velocity field $|v|\le q <c=1$ we have for
the dot product
\begin{equation*}
 |e\cdot v|\le q
\end{equation*}
so the field $z$ is well defined on the set $G$ and we have the
uniform estimate
\begin{equation*}
 0\le z\le \frac{1}{1-q}.
\end{equation*}

\begin{defin}[Fundamental fields]
\label{fundamental fields} The fields
\begin{equation*}
 \tau,\ r_{12},\ v,\ a,\ T,\ u,\ z,\ e
\end{equation*}
will be called the {\bf fundamental fields} associated with the
trajectory of the moving point mass.
\end{defin}


The fundamental fields are continuous on their respective domains.
This follows from the fact that composition of continuous
functions yields a continuous function. Thus all of them, for
sure, are continuous on the open set $G$ of points that do not lie
on the trajectory.

\bigskip

We would like to stress here that the fundamental fields depend on
the Lorentzian frame, in which we consider the trajectory. It is
important to find expressions involving fundamental fields that
yield fields invariant under Lorentzian transformations.

Lorentz \cite{lorentz1} and, independently, Einstein
\cite{einstein2a}, Part II, section 6, established that fields
satisfying Maxwell equations are invariant under Lorentzian
transformations.
\bigskip

Now consider the fields $E, B$ defined by means of fundamental
fields by the formulas
\begin{equation*}
 \begin{split}
E&=u^{2}e+u^{-1}D(u^2e)+D^2(e)\\
B&=e\times E
\end{split}
\end{equation*}
where $D$ denotes the partial derivative with respect to time
$\frac{\partial}{\partial t}.$

The pair $(E,B)$ of fields associated with a trajectory of a
moving point mass in a given Lorentzian frame satisfies homogenous
Maxwell equations. Therefore according to a result of Lorentz and
Einstein, the pair of fields $(E,B)$  is invariant under
Lorentzian transformations. The fields $E$ and $B,$ obtained
purely by mathematical construction, form parts of an
antisymmetric tensor of second rank.

Indeed, the pair of fields
$$E=(E_1,E_2,E_3)\qtext{and}B=(B_1,B_2,B_3)$$ is a part of an
antisymmetric tensor that in Lorentzian space-time $R^3\times R$
frame has the matrix that looks as follows
\begin{equation*}
    \begin{bmatrix}
      0 & +E_1 & +E_2 & +E_3 \\
      -E_1 & 0 & +B_3 & -B_2 \\
      -E_2 & -B_3 & 0 & +B_1 \\
      -E_3 & +B_2 & -B_1 & 0 \\
    \end{bmatrix}
\end{equation*}


When the point mass is in its rest frame the field $E$ differs
from Newton's gravity field just by a constant of proportionality.
The formula for the field $E$ amends the formula derived
heuristically by Feynman for a moving point charge. So it is
proper to call the field $E$ the {\bf Newton-Feynman field}
generated by the trajectory.

The following theorem permits us to analyze such fields $E$ and
$B$ by reducing the computations to simple algebraic expressions
involving the fundamental fields.

Introduce operators $D=\frac{\partial}{\partial t}$ and
$D_i=\frac{\partial}{\partial x_{i}}$ for $i=1,2,3$ and
$\nabla=(D_1,D_2,D_3).$ Observe that $\delta_i$ in the following
formulas denotes the i-th unit vector of the standard base in
$R^3$ that is $\delta_1=(1,0,0),$ $\delta_2=(0,1,0),$
$\delta_3=(0,0,1).$


\begin{thm}[Partial derivatives of fundamental fields]
Assume that in some Lorentzian frame we are given an   admissible
trajectory $t\mapsto r_2(t).$ Define the time derivative
$w(t)=\dot{r}_2(t).$ For partial derivatives with respect to
coordinates of the vector $r_1=y\in R^3$ we have the following
identities on the set $G$
\begin{eqnarray}
\label{DiT}       D_iT&=&ze_i  \qtext{where} v=\dot{r}_2\circ\tau,\\
\label{Diu}       D_iu&=&-zu^2e_i  ,   \\
\label{Div}          D_iv&=&-e_iza \qtext{where} a=\dot{w}\circ\tau,\\
\label{Di tau}    D_i\tau&=&-ze_i,\\
\label{Die}         D_ie&=& -uze_ie+u\delta_i+uze_iv \qtext{where} \delta_i=(\delta_{ij}),\\
\label{Diz}          D_i z &=& -z^3e_i \langle e,a \rangle
-uz^3e_i + uz^2e_i+uz^2v_i+uz^3e_i \langle v,v \rangle
\end{eqnarray}
and for the partial derivative with respect to time we have
\begin{eqnarray}
\label{DT}    DT&=&1-z,\\
\label{Du}    Du&=&zu^2-u^2,\\
\label{D tau}    D\tau&=&z,\\
\label{Dv}    Dv&=&za  \qtext{where} a=\dot{w}\circ\tau,\\
\label{De}    De&=&-u e+ u z e-u z v,\\
\label{Dz}    Dz&=&uz-2uz^2 +z^3 \langle e,a \rangle +uz^3-uz^3
\langle v,v \rangle .
\end{eqnarray}
Since the expression on the right side of each formula represents
a continuous function, the fundamental fields are at least of
class $C^1$ on the set $G.$ Moreover if the trajectory is of class
$C^\infty$ then also the fundamental fields are of class
$C^\infty$ on $G.$
\end{thm}


\section{Bogdan-Feynman Theorem for a moving point mass}
\bigskip

The following theorem is just a consequence that the wave emitted
from a trajectory of a point mass propagates in a Lorentzian frame
with the speed of light.  Please notice that the physical nature
of the fields is completely irrelevant.

The important fact is that we are working in a Lorentzian frame,
in which we are given a trajectory, on which the velocity and
acceleration are bounded. Such trajectory alone generates in a
unique way the system of the fundamental fields by means of which
we are able to construct fields that are preserved by Lorentzian
transformations.


We remind the reader that we are working in a fixed Lorentzian
frame. The position vector is denoted by $r_1\in R^3$ and time by
$t\in R.$ The position axes are oriented so as to form right
screw orientation. The units are selected so that the speed of
light $c=1.$

Partial derivatives with respect to coordinates of $r_1$ are
denoted by $D_1,\,D_2,\,D_3$ and with respect to time just by $D.$
The gradient differential operator is denoted by
$\nabla=(D_1,D_2,D_3)$ and the D'Alembertian operator by
$\Box^2=\nabla^2-D^2.$ The expression $\dotp{e}{v}$ denotes the
dot product in $R^3$ of the vectors involved.

\begin{thm}[Bogdan-Feynman Theorem]
\label{Bogdan-Feynman Theorem} Assume that in a given Lorentzian
frame the map $t\mapsto r_2(t)$ from $R$ to $R^3$ represents an
admissible trajectory of class $C^3.$ 
Assume that $G$ denotes the open set of points that do not lie on
the trajectory. All the following field equations are satisfied on
the entire set $G.$

Consider the pair of fields $E$ and $B$ over the set $G$ given by
the formulas
\begin{equation*}
     E=u^2e+u^{-1}D(u^2e)+D^2e\qtext{and}B=e\times E
\end{equation*}
where $u$ and $e$ represent fundamental fields (\ref{fundamental
fields}) associated with the trajectory $r_2(t).$

Then this pair of fields will satisfy the following homogenous
system of Maxwell equations
\begin{equation*}
    \begin{split}
    &\nabla\times E=-DB,\quad \nabla\cdot E=0,\\
    &\nabla\times B=+DE,\quad \nabla\cdot B=0,\\
    \end{split}
\end{equation*}
and the homogenous wave equations
\begin{equation*}
    \Box^2 E=0,\qquad \Box^2 B=0.
\end{equation*}

Moreover Li\'{e}nard-Wiechert potentials, expressed in terms of
the fundamental fields as  $A=uzv$ and $\phi=uz,$ satisfy the
homogenous system of wave equations with Lorentz gauge formula
\begin{equation*}
    \Box^2 A=0,\quad \Box^2\phi=0,\quad \nabla\cdot A+D\phi=0
\end{equation*}
and generate the fields $E$ and $B$ by the formulas

\begin{equation*}
\begin{split}
    E&=-\nabla \phi-DA\qtext{and}B=\nabla\times A.\\
\end{split}
\end{equation*}

Finally we have the following explicit formula for the field $E$
in terms of the fundamental fields
\begin{equation*}\label{formula4F-b}
\begin{split}
    E&= -uz^2a   +uz^3 \langle e,a \rangle e -uz^3 \langle e,a \rangle v \\
    &\quad  +u^2z^3e-u^2z^3 \langle v,v \rangle e-u^2z^3v+u^2z^3 \langle v,v \rangle v.\\
\end{split}
\end{equation*}
\end{thm}
\bigskip

For a proof of the above theorem see Bogdan \cite{bogdan65},
Theorem 11.1. As a consequence of the above theorem the components
of the quantities $E,$ $B,$ $A,$ and $\phi,$ since they satisfy
the homogenous wave equation $\Box^2 y=0,$
 propagate in the Lorentzian frame with velocity of light $c=1.$
\bigskip

\section{Relativistic equation  of motion for a spacecraft\\
in the joint gravitational field\\
 of the Earth and the Sun\\
 in a Lorentzian frame attached to the Earth}
\bigskip

Now for a body of rest mass $m_0$ moving under the influence of a
force $F$ from Newton-Einstein formula, time derivative of the
momentum is equal to force, we must have
\begin{equation*}
    \dot{p}=F
\end{equation*}
where $p=m_0\gamma\, v$ is the relativistic momentum and
$\gamma(v)=(1-|v|^2)^{-1/2}$. Computing the derivative of $p$ with
respect to time we get
\begin{equation*}
    \begin{split}
    \dot{p}&=m_0\,\gamma\, \dot{v}+m_0\,\dot{\gamma}\,v
    =m_0\,\gamma\, \dot{v}+m_0\,\gamma^3\,\dotp{v}{\dot{v}}\,v\\
    &
    =m_0\gamma\, ( \dot{v}+\gamma^2\,\dotp{v}{\dot{v}}\,v)=
    m_0\,\gamma(v)\,\Gamma(v)(\dot{v}),\\
    \end{split}
\end{equation*}
where $\Gamma(v)$ is the linear transformation of $R^3$ into $R^3$
given by the formula
\begin{equation}\label{Gamma}
 \Gamma(v)(h)=h+\gamma^2\dotp{v}{h}v\fa h\in R^3.
\end{equation}
Notice that for fixed velocity $v$ the transformation $\Gamma(v)$
represents a symmetric, positive definite transformation with
eigenvalues equal respectively to
\begin{equation*}
 (1+\gamma^2|v|^2),\ 1,\ 1.
\end{equation*}
Thus the inverse transformation $\hat{\Gamma}(v)$ exists and is
also symmetric. Since its eigenvalues are
\begin{equation*}
 (1+\gamma^2|v|^2)^{-1},\ 1,\ 1
\end{equation*}
and the norm of positive definite symmetric transformation is its
maximal eigenvalue, we must have
\begin{equation*}
 |\hat{\Gamma}(v)|=1\fa v\in R^3
\end{equation*}
and we have also
\begin{equation*}
 \hat{\Gamma}(v)\Gamma(v)=e\fa v\in R^3
\end{equation*}
where $e$ denotes here the identity transformation in the algebra
$Lin(R^3,R^3)$ that is $e(h)=h$ for all $h\in R^3.$

We shall call the function $v\mapsto \hat{\Gamma}(v)$ the {\bf
reciprocal to the function} $\Gamma(v),$ since
\begin{equation*}
 \hat{\Gamma}(v)=[\Gamma(v)]^{-1}\fa v\in R^3.
\end{equation*}
\bigskip
\section{The mystery force identified}
\bigskip

Thus the relativistic equation of motion of the spacecraft can be
written as
\begin{equation}\label{spacecraft motion eq}
 m_0\,\gamma(v)\Gamma(v)(\dot{v})=-m_e\,E_e-m_s\,E_s
\end{equation}
where $m_e$ and $m_s$ denote the rest mass of the Earth and of the
Sun, and where $E_e$ and $E_s$ represent the Newton-Feynman fields
generated by the Earth and the Sun respectively. The {\bf mystery
force} is $F_s=-m_s\,E_s$ and as Mr. Anderson suspected it is due
to Earth's rotation, or equivalently due to the fact that in a
Lorentzian frame attached to Earth the Sun moves around the Earth.
\bigskip

\section{Algorithm for computing the transformation Gamma hat}

\bigskip

The space $Lin(U,U)$ of linear bounded operators from a Banach
space $U$ into itself beside being closed under addition and
scalar multiplication is also closed under the composition of
operators: $P\circ Q\in Lin(U,U)$ for all $P,Q\in Lin(U,U).$ This
operation has the property that $|P\circ Q|\le |P|\,|Q|.$

\bigskip

\begin{defin}[Banach Algebras]
A Banach space $U$ is called a {\bf Banach algebra,} if it is
equipped with a bilinear operation $(u,w)\into u\,w,$ from the
product $U\times U$ into $U,$ that is associative:
$(u\,w)\,z=u\,(w\,z),$ and such that
\begin{equation*}
    |u\,w|\le |u|\,|w|\fa u,\,w\in U.
\end{equation*}
If in addition there is an element $e$ in $U$ such that
$e\,u=u\,e=u$ for all $u\in U,$ then such an algebra is called a
{\bf Banach algebra with unit.} In such a case an element $u$ is
called invertible if for some $w\in U,$ called the inverse of $u,$
we have
\begin{equation*}
    u\,w=w\,u=e.
\end{equation*}
The unit element and the inverse elements are unique. We denote
the inverse of $u$ by $u^{-1}.$
\end{defin}

\bigskip

The following proposition is very useful in numerical computations
of the inverse transformations. It represents a simple application
of the Banach fixed point theorem and it is essential in the
development of the theory presented in this note. It represents
Pro. 6.25 of Bogdan \cite{bogdan64}.
\bigskip
\begin{pro}[Inverse $(e-v)^{-1}$ exist if $|v|<1$]
\label{inverse (e-v)} Let $U$ be a Banach algebra with unit. Then
for every element $v\in U$ such that $|v|<1$ the inverse
$w=(e-v)^{-1}$ exists. It is a fixed point of the operator $f$
defined by the following formula
\begin{equation*}
    f(u)=e+vu\fa u\in U.
\end{equation*}
Moreover the sequence $w_n=f(w_{n-1})$ where $w_0=0$ is of the
form
\begin{equation*}
    w_n=e+v+v^2+\cdots+v^n=\sum_{0\le k\le n}v^k\fa n>0
\end{equation*}
and we have the following estimate for the distance of the fixed
point $w$ and the approximation $w_n$
\begin{equation*}
    |w-w_n|\le \frac{|v|^n}{1-|v|}\fa n>0,
\end{equation*}
and we also have an explicit formula for the inverse element $w$
as the sum of an absolutely convergent series
\begin{equation*}
    w=e+v+v^2+\cdots=\sum_{n\ge 0}v^n.
\end{equation*}
\end{pro}

\bigskip
\section{Returning to our problem at hand}
\bigskip

Notice first of all that the space $Lin(R^3,R^3)$ of all linear
transformations from $R^3$ into itself can be identified with the
space of $3\times 3$ matrices. If $e$ denotes the unit matrix let
$H(v)$ denote the transformation defined by the formula
\begin{equation*}
 H(v)(h)=-\gamma(v)^2\,\dotp{v}{h}\,v\fa h\in R^3.
\end{equation*}
The norm of this transformation is $\frac{|v|^2}{1-|v|^2}.$ So if
the velocity $|v|$ stays below $\sqrt{0.5}=0.707\ldots$ that is
below the $70\%$ of the speed of light the norm of the operator
$H$ will be less then $1.$ So we can use the above proposition to
compute the inverse transformation
\begin{equation*}
 \hat{\Gamma}(v)=(e-H(v))^{-1}
\end{equation*}
with any level of precision.

\bigskip

\section{Algorithm for computing relativistic trajectory of spacecraft}
\bigskip

So here we are. We can rewrite the equation of motion of the
spacecraft (\ref{spacecraft motion eq}) in the equivalent form as
\begin{equation*}
\begin{split}
 \dot{v}&=(m_0\gamma(v))^{-1}\hat{\Gamma}(v)(-m_e\,E_e-m_s\,E_s)=f(y,t,v),\\
 \dot{y}&= v,\\
\end{split}
\end{equation*}
and this system of differential equations we can solve, say, by
means of Runge-Kutta methods with any level of precision.

In closing I would like to thank Mr. Anderson from Jet Propulsion
Laboratory for putting this problem onto You Tube.

Taking this opportunity I would like to thank also to my former
colleagues, Donald Jezewski and Victor Bond, from Mission Planning
and Analysis Division of NASA's  Lyndon B. Johnson Space Center in
Houston, Texas, for introducing me to problems of celestial
mechanics.


\bigskip

\end{document}